\documentclass[aps,pra,reprint,english,twocolumn,showpacs,preprintnumbers,amsmath,amssymb,floatfix]{revtex4-1}

\usepackage[T1]{fontenc}
\usepackage[latin9]{inputenc}
\usepackage{graphicx}
\usepackage{amssymb}
\usepackage{babel}

\usepackage{babel}
\makeatother

\begin{document}

\title{Information Exchange via Surface Modified Resonance Energy Transfer}

\author{Mathias Bostr{\"o}m}
\email{Mathias.Bostrom@mse.kth.se}

\author{Clas Persson}

\affiliation{Department of Materials Science and Engineering, Royal Institute of Technology, SE-100 44 Stockholm, Sweden}

\affiliation{Department of Physics, University of Oslo, P. O. Box 1048 Blindern, NO-0316 Oslo, Norway}

\author{Dan Huang}
\affiliation{Department of Materials Science and Engineering, Royal Institute of Technology, SE-100 44 Stockholm, Sweden}

\author{Barry W. Ninham}

\affiliation{Department of Applied Mathematics, Australian National University, Canberra, Australia}

\author{Bo E. Sernelius}
\email{bos@ifm.liu.se}
\affiliation{Division of Theory and Modeling, Department of Physics, Chemistry and Biology, Link\"{o}ping University, SE-581 83 Link\"{o}ping, Sweden}

\begin{abstract}
The theory is presented for resonance interaction between two atoms in an excited configuration: one atom, the "receptor" of information (i.e. energy), adsorbed on a phospholipid surface and the other atom, the "emitter" of information (i.e. energy), a long distance away.  The dielectric function for a specific phospholipid membrane is obtained from density functional theory calculations. We present numerical results comparing the range and magnitude of non-specific Casimir-Polder interactions with the much more long-ranged, and highly specific, resonance interaction.  A study of the resonance interaction with one or both atoms adsorbed on a phospholipid membrane surface reveals a possibility to have a cross over from attraction to repulsion or from repulsion to attraction at separations  between  receptor and emitter atoms exceeding several hundred {\AA}ngstr{\"o}ms. The energy transfer and the observed transitions in the sign of the interaction energies near surfaces provide potential new ways to start recognition processes in biological systems. 

\end{abstract}


\maketitle

\section{Introduction}

F\"{o}rster energy transfer\,\cite{Forster} goes to the very heart of biology and biophysics.  It underlies photosynthesis, artificial light harvesting, and fluorescent light-emitting devices\,\cite{Groendelle1,Bergstrom1,Lane}. Resonant energy transfer was discovered experimentally by Cario and Franck in 1923\,\cite{Cario}. Considering its importance it is no surprise that many experiments have been carried out to investigate the properties of F\"{o}rster energy transfer, e.g. effects due to confinement\,\cite{Andrew}, solvent and temperature\,\cite{Renge}.  The dipole-dipole mechanism has furthermore been proposed as a way to create entangled states for quantum logic using both molecules\,\cite{Brennen} and quantum dots\,\cite{Matsueda}. 
Excited state resonance interaction between two atoms is of much longer range than Casimir-Polder interaction between two ground state atoms\,\cite{Sherkunov,Stephen,McLachlan,Casimir}.  Interestingly, the  retarded resonance interaction can  influence  the binding energy of atom pairs\,\cite{Jones,McAlexander}. 
In pores and near surfaces these interactions may be modified.\,\cite{BosEPL2013,BosPRA2013}  Experimental\,\cite{Hopmeier} and theoretical\,\cite{Agarwal}  evidences have been found for the enhancement of dipole-dipole interactions in microcavities. We have recently predicted a way by which very large molecules may form by resonance interaction between two identical atoms in a narrow cavity\,\cite{BostrPRA2012}.  A  problem of significance to catalyzis is how breakage and formation  of molecules may occur near surfaces\,\cite{Blum, darling}. 
Interactions driven by electromagnetic fluctuation forces, and real photon exchange between molecules, are at the foundations of much of physical and biological sciences. They determine the properties of condensed matter, chemical reactions, recognition, catalyzis, and self-organization in nanotechnologies and biology. 

We will, as a motivation for a study on information exchange via resonance interaction between atoms near surfaces, first outline an intriguing, but speculative, idea by B. W. Ninham\,\cite{NinBost}  on pheromone action via photon transfer\,\cite{Forster}. The basic theory of F\"{o}rster energy transfer is discussed. We then proceed to present the theory of resonance interaction between an atom adsorbed on a surface ("receptor" of information, i.e. photon energy) and an atom far away from the surface ("emitter" of information). Once the information (i.e. energy) has been received by the receptor it can be transferred to its final destination via standard biochemical transfer mechanisms. There may be biological systems where this model can be used as an approximation. We consider as an illustration the case of two helium atoms near a phospholipid membrane and demonstrate that resonance interaction is both larger in magnitude and of much longer range than the Casimir-Polder (van der Waals) interaction. We end this paper with our main conclusions.

\section{A Biolgical Motivation for Studying Excited State Atom-Atom Interactions Near Surfaces }

We recapitulate  in this section an intriguing, and speculative, idea by  Bostr{\"o}m and Ninham\,\cite{NinBost} on pheromone action taking place via photon transfer.  While the sequence of biochemical and neurophysiological events consequent on recognition of a pheromone is well understood, the recognition process is not. A new, testable, mechanism was proposed\,\cite{NinBost} to explain how highly specific pheromone molecules transfer information and activate receptor proteins within antennae or combs of insects. The primary purpose of this article is to present the theory for excited state atom-atom interaction with one atom adsorbed to a surface and the other far way. The secondary purpose of this article is aiming to stimulate a healthy debate about the origin of the pheromone recognition process, as well as stimulating further experiments.

Pheromones play a vital role in communication between insects in finding food as well as recognition and location of a mate. Sex pheromones are emitted by the female and recognized by the male at extremely low concentrations.\,\cite{Carde} The male then flies up wind, following the trail of the pheromone plume, and eventually reaches the female. All pheromones are simple, non-ionic, non-reactive hydrophobic molecules. Typically they are short-chained molecules, e.g. C12 hydrocarbons, with a simple terminal or medial group, such as aldehydes, alcohols and acetates. They differ, in the nature of this group, and positions of occasional double bonds. The emitted pheromone plume from the female often consists of two or more compounds. For the pheromonal communication system to be effective, the male has to be able to single out the specific pheromone blend from other blends emitted by different species. Odor discrimination is accomplished by the sensitive olfactory system, which is primarily located in the insect antennae.\,\cite{Carde,Kanaujia}

Recognition of the pheromone molecule is supposed to proceed as follows\,\cite{Carde,Ninhb}: The (hydrophobic) pheromone is physisorbed via van der Waals forces\,\cite{Ninhb} on the solid hydrophobic antennae. The adsorbed pheromone then diffuses along the surface until it finds a molecular sized pore tubule containing an aqueous sensillar lymph protecting the sensitive olfactory neurons. The sensillar lymph contains a highly abundant protein that specifically binds to pheromones, thereby facilitating the transport of the hydrophobic odor molecules through the aqueous lymph to the receptor at the dendritic membrane. Upon reaching the receptor site, either the pheromone molecule alone or the pheromone-protein complex activates the receptor, which give rise to action potentials that in turn elicit a behavioral response. For the olfactory system to be effective, it is of great importance that the pheromone molecules are rapidly eliminated to maintain high sensitivity towards additional pheromones.\,\cite{Carde}   It has been shown that little desorption takes place on the antennae which indicates that most of the adsorbed pheromones enter the sensillum as well as that the desorption process can be ruled out as a mechanism of removal and inactivation of the odor molecules.\,\cite{Kanaujia} This also implies that all the molecules adsorbed on the antennae are degraded in the sensillar lymph. It has been suggested that the pheromone binding proteins as well as certain enzymes might be involved in the inactivation and degradation of the pheromone molecules.\,\cite{Carde} 

There are evidences to doubt the standard hypothesis. There is nothing specifically different in the visible or ultra-violet spectrum of all pheromone molecules that determine the nature of the van der Waals forces. There can be no discrimination in adsorption, surface diffusion or protein-pheromone interaction. Further, every other hydrophobic molecule in the atmosphere, present in billions of times larger concentrations would also bind to the antenna with the same van der Waals forces. The antenna of the unfortunate male insect would be coated with a thick film of all these other molecules. This would effectively prevent detection of the relevant pheromones emitted by the female insect.

There is an additional well-known objection. The proposed van der Waals interaction is far too weak and far too short-ranged to account for recognition of the extremely low concentrations of pheromones emitted by the female.\,\cite{Karg} The mass balances simply do not add up and are missing factors of $10^6$ or more. That conclusion from numerous experiments is despite the fact that only a few receptor molecules need to be activated to trigger a response. Some other much longer ranged specific communication must be involved.

In seeking other possible explanations one may note that no chemical reaction between pheromone and receptor protein seems to be involved.\,\cite{Kanaujia} Since van der Waals forces, due to molecular polarizabilities in the visible and ultraviolet are non-specific and too short ranged, we consider the infrared region. Here certainly the pheromone molecules differ substantially. In the infrared the response functions are indeed highly specific. Ninham\,\cite{NinBost,Ninhb} then hypothesized that a pheromone is emitted by the female in a long-lived metastable excited state. The pheromone may also be excited into this long-lived metastable excited state by the infrared (IR) spectrum of sunlight. The vibration modes of the protein receptor molecule in different conformations, again in the infrared, are also highly specific. If the energy of one of these excited states coincides with that of the metastable pheromone, then energy transfer can take place through photon exchange via the quantum mechanical resonance energy mechanism.\,\cite{Bostrom1,BostrPRA2012,Sherkunov} These interactions between excited and ground state molecules are stronger, of longer range, highly directional, and most importantly highly specific. At distances large compared to molecular scales of van der Waals interaction we can envisage a process whereby the pheromone molecule in its metastable excited state identifies an identical resonance frequency in the protein receptor and transfers a photon of precisely the right energy to induce the required conformational change. The conformation change so induced allows the receptor protein to be released.  Pheromones certainly will have resonances in the infrared energy spectrum. Some must coincide with those in the IR spectra of candles. Which might explain the fatal attraction of moths to candle flames and the frequently observed mass immolation of insects attracted to Australian bush fires. Further, insects are not too much engaged in mating when the atmosphere is moist---water is a very strong infrared adsorber and likely to de-excite the pheromone molecules. As further hint that such a physical, as opposed to chemical mechanism might be involved Bostr{\"o}m and Ninham noted too that fireflies evidently communicate in the visible region, but the emission spectrum must contain a weak contribution from the infrared.

\begin{figure}
\includegraphics[width=6cm]{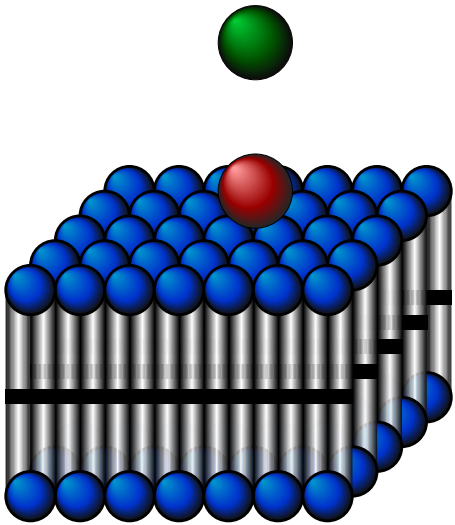}
\caption{(Color online) Carton figure of a receptor atom bound at the surface of a phospholipid membrane interacting with a second information carrying atom away from the surface. Information is transfered via photon exchange. }
\label{figu1}
\end{figure}

The principal objection to this hypothesis is whether or not a pheromone molecule can be emitted in a long-lived metastable state. A relevant analytic treatment of a model system demonstrated that for a chain of coupled oscillators with a defect a long lived eigenstate can emerge from the band (continuum of excitation frequencies).\,\cite{NinhA} The pheromone molecule can be modeled as a system of identical (hydrocarbon) harmonic oscillators with an occasional different spring constant (terminal and at double bond on the chain). So certainly such states exist. This long-lived excited state would be specific to each pheromone. 
The simplest answer to this objection is to test the hypothesis by experiment. Isolation of male insects in a closed container with an infrared window and exposure to a very weak tunable infrared laser source either evinces a response or does not. Estimates of an appropriate range of frequencies for particular pheromone molecules might be obtained via quantum chemistry calculations. 

We have recently reviewed the resonance phenomenon that is central to the Ninham hypothesis for the pheromone action.\,\cite{Bostrom1,BostrPRA2012} We demonstrated that the previous quantum perturbative treatment for the resonance interaction is incorrect in the long-range retarded limit. It is now understood that correct results require a self-consistent treatment.  We define the distance between two atoms to be $\rho$ [\AA]. When the two oscillators come close together the resonance interaction is proportional to $ \rho^{-3}$, for larger separations retardation becomes important and the interaction decays more rapidly proportional to $\rho^{-4}$. Near a phospholipid surface the excited and ground state atom-atom interactions are strongly modified. We show our model system in Fig.\,\ref{figu1}. As we will see for the resonance interaction between two atoms near a biological surface there can even be a transition from attraction to repulsion or from repulsion to attraction depending on the orientation of the atoms relative to the phospholipid membrane. 

What is of prime importance for the present description of pheromone action is that the resonance interaction is many orders of magnitude larger and of much longer range than the van der Waals interaction between the pheromone molecule and the receptor protein (which asymptotically vanishes as $\rho^{-7}$). It is also of much longer range than the van der Waals interaction between pheromone molecule and the cylindrical antenna (which asymptotically vanishes as $\rho^{-6}$). Close enough to the cylindrical surface the resonance interaction has the same power-law and similar magnitude as the van der Waals interaction towards the interface. However, the resonance interaction is highly directional towards the pore tubule or more specifically towards the receptor protein. More important, this interaction does not require that the pheromone actually come into physical contact to activate the receptor protein. This also offers the obvious explanation to why the cell response on termination of pheromone exposure is much too fast to be explained by ordinary diffusion transport.\,\cite{Kanaujia}  The response time is obviously much faster within the resonant photon transfer model. There is within the Ninham hypothesis no response at all when there is no excited pheromones around.

\section{Self-Consistent Theory for Resonance Interaction and F\"{o}rster Transfer}

Before we derive the resonance interaction energy between two identical atoms in an excited configuration we first define what we here mean by transfer rate. In the strong-coupling limit one may define the rate of ``fast'' transfer between two identical atoms, one in the ground state and the other in an excited state, as\,\cite{Forster}:
\begin{equation} 
n \approx 2  |U|/(\pi \hbar),
\label{Eq1}
\end{equation}
where $U$ is the resonance energy and $\hbar$ is Planck's constant. It should be stressed that the energy transfer rate depends on the underlying Green's function and is related to the resonance interaction energy. F\"{o}rster demonstrated how the transfer rate of both strongly and weakly coupled atoms or molecules can be treated within the same formalism\,\cite{Forster}. Between two weakly  interacting molecules, that may in general be different, there is enough that there is an overlap of the energy-bands to have energy transfer. Application of time-dependent perturbation theory gives the following approximation (Fermi golden rule rate) for this ``slow'' transfer rate\cite{Forster}:
\begin{equation}
n \approx 2\pi |U|^2 \delta/\hbar,
\label{Eq2}
\end{equation}
where $\delta$ is the ``density of final states''  (related to the spread in the energy of the optical band associated with slow energy transfer\,\cite{Forster}).

Writing up the equations of motion for the excited system it is straightforward to derive the zero temperature Green function for two identical (and isotropic) atoms\,\cite{McLachlan,Bostrom1}. The transition moments of the two atoms are assumed to be parallel and making an angle with the axis that joins the two atoms. The extension to consider two different atoms is trivial and the resonance frequencies ($\omega_r$) of the system are given by the following equation:
\begin{equation}
1-\alpha(1|\omega) \alpha(2|\omega) T(\rho|\omega)^2=0,
\label{Eq3}
\end{equation} 
where $\alpha(j|\omega)$ is the polarizability of atom $j$ and $T(\rho|\omega)$ is the susceptibility tensor. The tensor elements corresponding to interactions in air are given in the literature.\,\cite{Bostrom1} In the case of two identical atoms the above resonance condition can be separated in one anti-symmetric and one symmetric part. The excited symmetric state has a much shorter life time than the excited anti-symmetric state leading to the well known fact that the system can be trapped in an excited antisymmetric state.  The resonance interaction energy of this antisymmetric state is,
\begin{equation}
\label{Eq5}
U(\rho)= \hbar [\omega_{r} (\rho)-\omega_{r} (\infty)].
\end{equation}
Since the relevant solution of  Eq.\,(\ref{Eq3}) really is the pole of the antisymmetric part of the underlying Green function we can in a standard way\,\cite{Lifshitz,Mahanty2,Sernelius} deform a contour of integration around this pole to obtain a both simple and exact expression for the resonance interaction energy,
\begin{eqnarray}
U(\rho) = (\hbar/ \pi) \int_0^\infty d \omega \ln[1+\alpha(1|i \omega) T(\rho|i \omega)].
\label{Eq6}
\end{eqnarray}

In the weak coupling limit the atoms can be either of the same kind or different. Here for the sake of demonstration we only consider two identical atoms. In principle the transition rate should be evaluated as an integration over the overlap of energy levels of the two atoms. However, here we make use of the approximate Eq.\,(\ref{Eq2}). In this way we focus only on the part of the transfer rate that has been evaluated incorrectly, i.e. the coupled dipole-dipole interaction (which here is $n \propto U^2$). In free space we find that the non-retarded transfer rate is $\propto \rho^{-6}$ which is the F\"{o}rster transfer rate. At zero temperature the long-range retarded asymptote decays more rapidly as $ \rho^{-8}$. 
Finite temperature effects can be easily dealt with as for the corresponding ground state problem\,\cite{Daicic,Daicic2,MJBPRA}. In fact as for the interaction between two ground state atoms the correct long-range interaction can only be found when finite temperature is accounted for\,\cite{Daicic,Daicic2}. 
Expressed in terms of the classical Bohr radius ($a_0$) and the first excited state of a real hydrogen atom the long-ranged purely classical result is $k_B T a_0^3/\rho^3$.  The long range F\"{o}rster transfer rate is therefore proportional to ${k_B}^2 T^2/\rho^6$. This manifestation of the correspondence principle is very similar in nature to the result obtained for the retarded van der Waals interaction between two ground-state atoms\,\cite{Daicic}. It is not only the correct result at high enough temperatures, but also at any finite temperature for large enough separations. The change in power-law at large separations has usually been interpreted as being simply due to the finite velocity of light. However, the long-range interaction at finite temperatures between two atoms is independent of the velocity of light. This demonstrates that there is more to it than a simple loss of inter-correlation due to the finite velocity of light. As pointed out by Wennerstr\"{o}m, Daicic, and Ninham\,\cite{Daicic2} the quantum nature of light must be important to the softening of the interaction potential.

\section{Flaws in Perturbative QED}

The incorrect results in free space, obtained from perturbative QED, came out  from our formalism if certain simplifying assumptions were made.\,\cite{Bostrom1}  The pole of the antisymmetric polarizability, i.e. the resonance frequency, is close to the oscillator frequency ($\omega_j$) so one should think that replacing $T(\rho|\omega)$ with $T(\rho|\omega_j)$, and dropping the dissipation frequency (i.e. the lifetime), should be a valid approximation. Indeed, this is a frequently used approximation\,\cite{McLachlan,Brennen}. The resonance frequency is within this approximation,
\begin{equation}
\omega_r\approx\omega_j \sqrt{1+\alpha(0) T(\rho|\omega_j)}\approx \omega_j+\omega_j \alpha(0) T(\rho|\omega_j)/2.
\label{Eq12}
\end{equation} 

Using the definitions of the oscillator strength and the static polarizability\,\cite{McLachlan} we can write the interaction energy as,
 \begin{equation}
U(\rho)=p^2 T(\rho|\omega_j),
\label{Eq13}
\end{equation}
where $p$ is the magnitude of the transition dipole moment. This is the classical text-book result obtained in various derivations distributed in the literature over a period of almost 40 years.  Another way to derive the same result is to first calculate the displacement vector field of the non-interacting excited molecules. The result is then obtained by calculating the energy of the dipole of the other molecule in this field\,\cite{Power,Power2}. The main difference compared to the derivation in the previous section is obviously that the coupling of the system has been totally neglected. 

In the strong-coupling limit the retarded asymptotic transfer rate is $n \propto \rho^{-1} cos(\omega_j \rho/c)$. In the weak-coupling limit the transition rate is as before obtained using Fermi golden rule. In the ``standard approach''\,\cite{Andrews} the real part of the field susceptibility is used. The averaged isotropic retarded transfer rate becomes proportional to $\rho^{-2} cos^2(\omega_j \rho/c)$. The oscillating transfer rate has been recognized as being incorrect and a way to avoid these oscillations is to use the complex field susceptibility. After averaging isotropically the following result is obtained\,\cite{Avery,Power,Power2,Andrews},
\begin{equation}
n \propto [3+(\omega_j \rho/c)^2+(\omega_j \rho/c)^4]/\rho^6.
\label{Eq14}
\end{equation}
In the non-retarded limit the result is identical to what we derived in the previous section. However, it is totally different in the retarded limit. The argument used to support the $\rho^{-2}$-dependence has been that it supposedly corresponds to real photon exchange. The $\rho^{-2}$-dependence is characteristic of a classical spherical wave. This separation dependence results in unphysical infinities for an increasingly large system of molecules. 
A way to avoid this has been to take into account the influence of the back-ground medium\,\cite{Juze}. In this way the interaction in condensed matter become modulated by an exponentially decaying factor. However, the theory used is on exactly the same inadequate level of approximation as previous work in free space. 
The reason why these perturbative (both QED and semiclassical) approaches fail is that they do not properly take into account that this is a dynamical system where the two atoms are coupled through the field.

\section{Excited State Resonance Interactions Between a Surface Bound Receptor Atom and a Free Emitter Atom}

The resonance interaction is strongly modified near surfaces and in pores. We have for example demonstrated the occurance of bound state atom pairs in narrow channels due to surface modulated resonance interaction.\,\cite{BostrPRA2012}
As a way of demonstrating that resonance interaction is much longer range than the van der Waals (Casimir-Polder) interaction we consider a simplified model system. We explore a system with a helium atom bound to a phospholipid surface and a second atom either also bound to the surface or above the first some distance from the surface. 

We have obtained a both simple and exact expression for the finite temperature resonance interaction energy between two identical polarizable particles excited in the {\it j}-direction ({\it j = x, y, or z}) placed in the {\it xz}-plane ({\it z} is in the direction away from the surface),

\begin{equation}
{U_j}(\rho ) = 2{k_B}T\sum\limits_{n = 0}^{\infty}{'}  {\ln [1 + \alpha (i{\xi _n}){T_{jj}}(\rho |i{\xi _n})]} .
\label{Eq15}
\end{equation}
where $k_B$ is the Boltzmann constant, $T$ is the temperature, and the prime indicates that the $n=0$ term should be divided by 2. Furthermore $\alpha(i\xi_n )$ is the atomic polarizability at the Matsubara frequencies  $\xi_n=2 \pi k_B T n/\hbar$.\,\cite{Bostrom1,Lifshitz,Sernelius} Here we consider a system at room temperature, so the temperature is T=300K. This will be compared with the ground state Casimir-Polder interaction which at large separations goes as

\begin{equation}
\begin{array}{*{20}{l}}
{{U_{CP}}(\rho ) \approx  -k_B T \sum\limits_{n=0}^{\infty}{' }  {\alpha ^2}(i \xi_n )}\\
{\quad \quad \quad \quad  \times \left\{ {\left[ {\sum\limits_{j = x,y,z} {T_{jj}^2} (\rho |i \xi_n )} \right] - 2T_{xz}^2(\rho |i \xi_n )} \right\}}
\end{array}
\label{Eq16}
\end{equation}

The susceptibility tensor is a sum of terms from free space susceptibility  ($ T_{jj}^0$), plus terms from surface corrections due to the presence of a surface. The susceptibility tensor elements are given in the literature.\,\cite{Buhmann1,Buhmann2,BosEPL2013} We have recently developed the theory for resonance interaction between two atoms in a narrow planar slit\,\cite{BostrPRA2012} and near a water drop.\,\cite{BosEPL2013} Here we will exploit the theory with arbitrary orientations of the two atoms near a phospholipid surface.

\section{Modeling the dielectric properties of a phospholipid membrane}

\begin{figure}
\includegraphics[width=8cm]{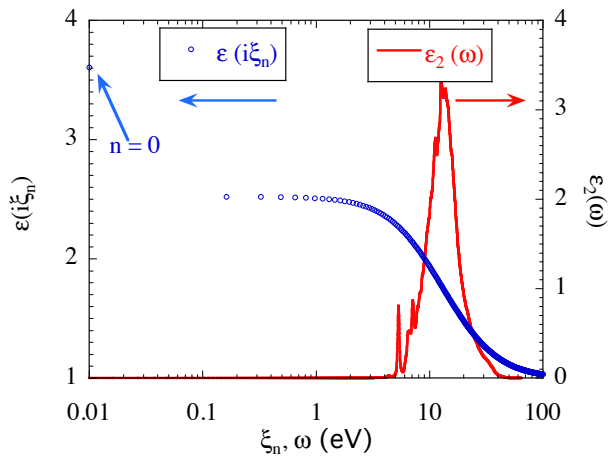}
\caption{(Color online) The dielectric function (left vertical axis) of a phospholipid membrane for discrete imaginary frequencies (at T=300K). Note that we have placed the $n = 0$ value at the left vertical axis. Also shown is  the corresponding imaginary part of the dielectric function for real frequencies (right vertical axis).}
\label{figu2}
\end{figure}

Our first-principles calculations are carried out within the density functional theory (DFT) as implemented in the Vienna Ab-initio Simulation Package (VASP).\,\cite{Kresse1,Kresse2} For the exchange correlation functional, the Perdew-Burke-Ernzerhof (PBE) generalized gradient approximation (GGA) method\,\cite{Perdew} is employed. An energy cutoff of 500 eV and a special k-point sampling over a $2\times4\times4$ Monkhorst-Pack mesh\,\cite{Monk} are used. The adopted phospholipids structure is deoxylysophosphatidylcholine monohydrate (3-dodecanoyl-propandiol-1-phosphorylcholine $\cdot$ H$_2$O) of which the lattice constants and the atomic positions except for hydrogen atoms are supplied through single-crystal analysis by Hauser.\,\cite{Hauser} We add the hydrogen atoms in our calculation model and after that we get the final structure by relaxing them. The imaginary part of the dielectric function is calculated by an independent-particle approximation. Thereafter, the spectra on the imaginary frequency axis are obtained via the Kramers-Kronig relation. The band gap correction is corrected by a simple scissors operator, in which the corrected value (1.68 eV) is obtained from the $\Gamma$ point band-gap difference between HSE06\,\cite{Krukau} and GGA-PBE. The static ion-clamped dielectric matrix is calculated by using the density functional perturbation theory.\,\cite{Gajdos}
The imaginary part of the dielectric function as well as the dielectric function for discrete imaginary frequencies are shown in Fig.\,\ref{figu2}.

\section{Numerical Results}

\begin{figure}
\includegraphics[width=8cm]{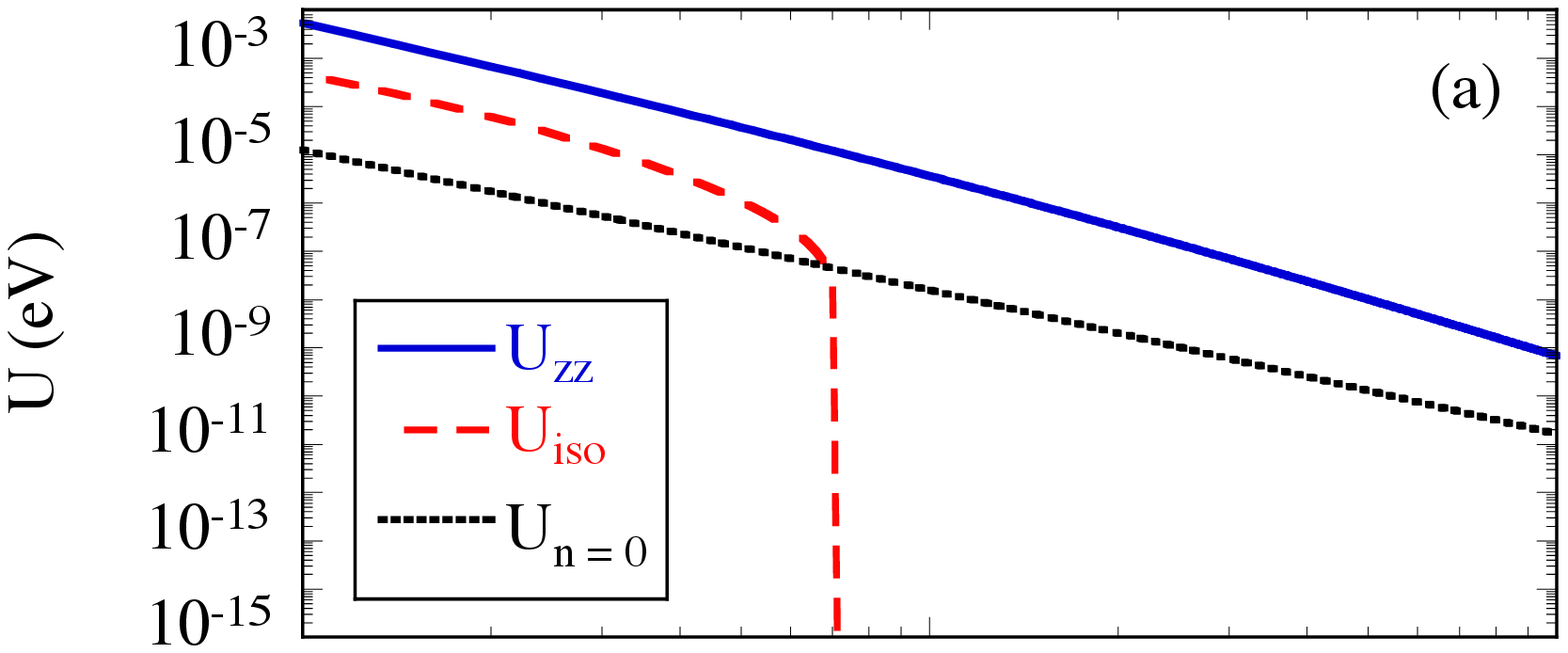}
\includegraphics[width=8cm]{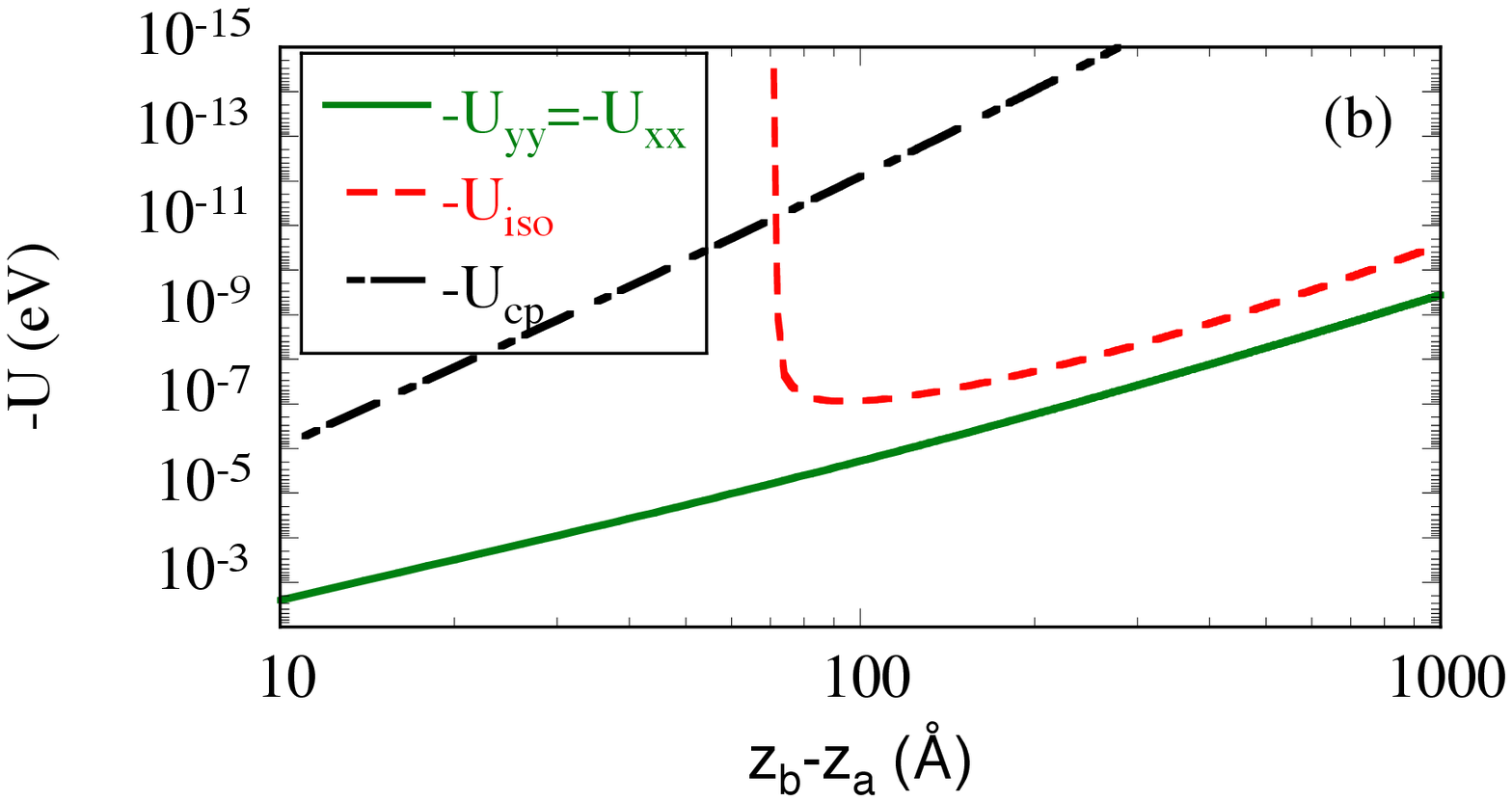}
\caption{(Color online) The  full resonance interaction energy between two helium  atoms with the {\it x}-branch, {\it y}-branch, and {\it z}-branch excited, and with all three excited (isotropic excitation) in an anti-symmetric excited state  as functions of separation, $\rho=z_b-z_a$. For comparison we also show the zero frequency contribution to the resonance interaction with $z$-branch excited. We consider the case with one of the two atoms adsorbed at a phospholipid surface ($z_a$=2 \AA)  and the other above it  far from the surface. We show also the corresponding result for the Casimir-Polder interaction between two ground state helium atoms.}
\label{figu3}
\end{figure}

\begin{figure}
\includegraphics[width=8cm]{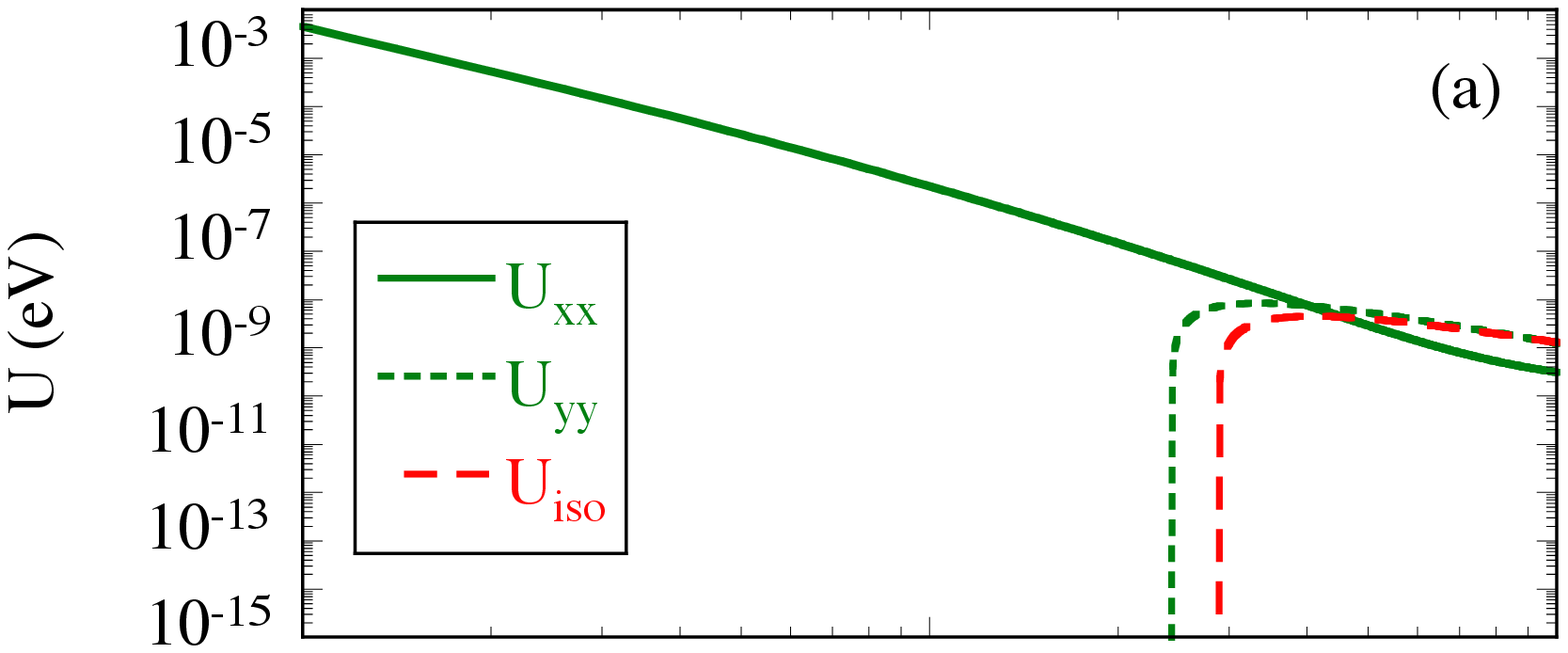}
\includegraphics[width=8cm]{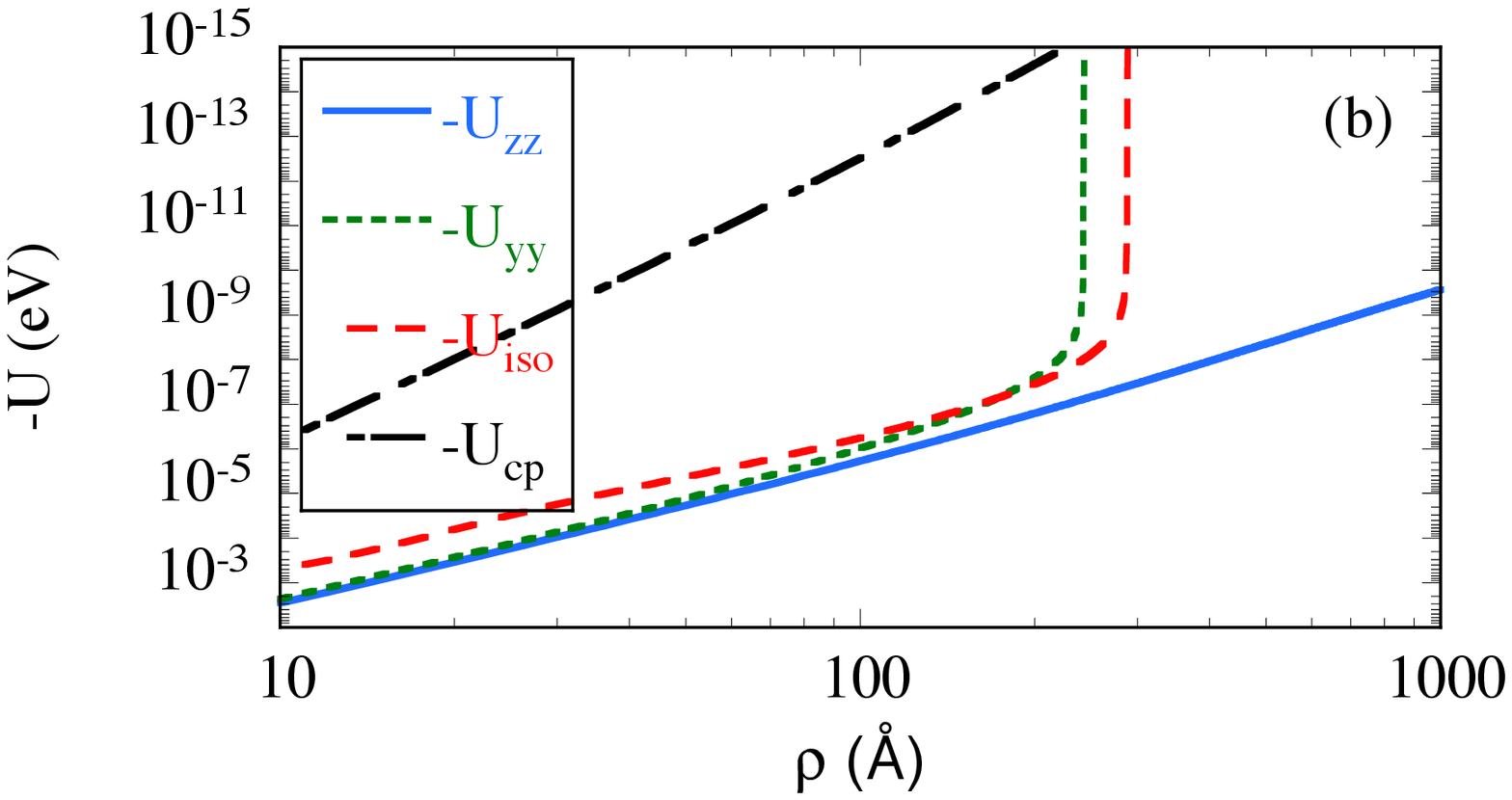}
\caption{(Color online) The resonance interaction with the $x$-branch, $y$-branch, and $z$-branch excited, and with all three excited (isotropic excitation) in an anti-symmetric excited state  as functions of separation for two helium atoms adsorbed on a phospholipid surface, $\rho=x$  ($z_a$=$z_b$=2 \AA).
We show also the corresponding result for the Casimir-Polder interaction between two ground state helium atoms. }
\label{figu4}
\end{figure}

We present results for the information exchange via energy transfer between receptor and emitter atom.  The result is shown  in Fig.\,\ref{figu3}  for the resonance interaction between a helium atom bound to a phospholipid surface interacting with an excited helium atom far from the surface ($x=0$ \AA).  The corresponding results for two helium atoms adsorbed on the phospholipid surface is shown in Fig.\,\ref{figu4}. 

We compare in Figs.\,\ref{figu3} and\,\ref{figu4}  the resonance interaction and the ground state Casimir-Polder interaction between two helium atoms near a phospholipid surface. The resonance interaction is of much longer range and larger in magnitude as compared to the ground state Casimir-Polder interaction.  The Casimir-Polder interaction is for all separations attractive and decreases monotonically towards zero. One should note that the resonance interaction between excited state atom pairs in free space is also monotonically decaying.\,\cite{Bostrom1} In contrast the resonance interaction with one or both atoms adsorbed on a surface reveals a cross over from attraction to repulsion or from repulsion to attraction at quite large separations between  receptor and emitter atom (around 60-300 \AA). These interaction energies and the corresponding energy transfer may provide the information exchange between the two atomic systems required to start the recognition process.

\section{Conclusions}

A  theory has been presented for resonance interaction between one ground state atom bound to a surface and a second excited state atom far away. We have demonstrated that the resonance interaction between two atoms in an antisymmetric state is of much longer range and larger in magnitude as compared to the Casimir-Polder interaction.We conclude that the resonance interaction is directional, of long range, and highly specific.  A study on the resonance interaction with one or both atoms adsorbed on a phospholipid membrane surface reveals a possibility to have a cross over from attraction to repulsion or from repulsion to attraction at separations  between  receptor and emitter atom exceeding several hundred {\AA}ngstr{\"o}ms. These changes in the sign of the interaction energies  near surfaces, and the corresponding changes on the energy transfer, may provide the required information exchange between the two atomic systems needed to start recognition processes.   Correctly evaluated resonance interaction provides some new understanding of the  long range of information  exchange via surface modified resonance energy transfer. 

At the present time there exists no explanation for the highly specific mechanism of pheromone action.  The initiating recognition event, assigned to van der Waals forces can be  ruled out.\,\cite{NinBost} The traditional mechanism is not discriminating and even if it were, the available pheromones are insufficient by at least a factor of millions. Since no chemistry is involved in the recognition the only possible source of specificity lies in the infrared spectrum of the pheromones. If these are presented in a metastable excited state, then communication and information transfer to the receptor protein that induces the necessary conformation change is accessible via the long-range specific directional resonance interaction. One should be clear about one important thing: the Ninham hypothesis for recognition between biosystems may be too simplistic to fully explain the example of insect pheromone action. But we argue there can be a contribution from this effect  in many systems relevant for recognition and information exchange in biology and nanotechnology.  One such example which requires the theory of information exchange extended to the case of unequal molecules is the catalyzis problems in biology, e.g. ATP-ADP reactions catalyzed at a cell membrane surface. We will return soon to the issue of energy transfer between unequal molecules in solution, free space and near interfaces.

MB and CP acknowledge support from VR (Contract No. C0485101) and STEM (Contract No. 34138-1). B.E.S. acknowledges financial support from VR (Contract No. 70529001).
Drew F. Parsons (Australian National University) is gratefully acknowledged for fruitful discussions and for providing the atomic polarizability.

\end{document}